\documentclass[english,aps,a4paper,groupedaddress,preprintnumbers,floatfix,nofootinbib,showpacs,twocolumn]{revtex4-1}

\usepackage[utf8]{inputenc}
\usepackage{graphicx}
\usepackage{enumerate}
\usepackage{amsmath}
\usepackage{slashed}
\usepackage{amssymb}
\usepackage{float}
\usepackage{amsfonts}
\usepackage{xcolor}
\usepackage{url}
\usepackage{hyperref}
\usepackage{subfig}
\usepackage{booktabs}
\usepackage{caption}
\usepackage{ragged2e}
\DeclareCaptionJustification{justified}{\justifying}
\usepackage[none]{hyphenat}
\hypersetup{colorlinks=true,linkcolor=redLinks,citecolor=greenLinks,
urlcolor=redLinks,
pdfborder={0 0 1}}
\hypersetup{
    bookmarks=true,         
    unicode=false,          
    pdftoolbar=true,        
    pdfmenubar=true,        
    pdffitwindow=false,     
    pdfstartview={FitH},    
    pdftitle={My title},    
    pdfauthor={Author},     
    pdfsubject={Subject},   
    pdfcreator={Creator},   
    pdfproducer={Producer}, 
    pdfkeywords={keyword1, key2, key3}, 
    pdfnewwindow=true,      
    colorlinks=false,       
    linkcolor=red,          
    citecolor=green,        
    filecolor=magenta,      
    urlcolor=cyan           
}

\begin{document}
\title{Constraints on neutrino electric millicharge from experiments of elastic neutrino-electron interaction and future experimental proposals involving coherent elastic neutrino-nucleus scattering}
\author{A. Parada\;}
\email{alexander.parada00@usc.edu.co} 
\affiliation{Facultad de Ciencias B\'asicas, Universidad Santiago de Cali, Campus Pampalinda, Calle 5 No.62-00, C\'odigo Postal 760035, Santiago de Cali, Colombia}

\sloppy

\begin{abstract}
  \noindent
  In several extensions of the Standard Model of Particle Physics (SMPP), the neutrinos acquire electromagnetic properties such as the electric millicharge. Theoretical and experimental bounds have been reported in the literature for this parameter. In this work, we first carried out a statistical analysis by using data from reactor neutrino experiments, which include elastic neutrino-electron scattering (ENES) processes, in order to obtain both individual and combined limits on the neutrino electric millicharge (NEM). Then we performed a similar calculation to show a estimate of the sensitivity of future experiments of reactor neutrinos to the NEM, by involving coherent elastic neutrino-nucleus scattering (CENNS). In the first case, the constraints achieved from the combination of several experiments are $-1.1\times 10^{-12}e < q_{\nu} < 9.3\times 10^{-13}e$ ($90\%$ C.L.), and in the second scenario we obtained the bounds $-1.8\times 10^{-14}e < q_{\nu} < 1.8\times 10^{-14}e$ ($90\%$ C.L.). As we will show here, these combined analyses of different experimental data can lead to stronger constraints than those based on individual analysis. Where CENNS interactions would stand out as an important alternative to improve the current limits on NEM.

\end{abstract}


\maketitle

\section{Introduction}
\noindent In the SMPP, the neutrinos are massless, electrically neutral, and only interact weakly with leptons and quarks. Nevertheless, the neutrino oscillation experiments show that neutrinos have mass and are also mixed \cite{Ahmad:2001an,Fukuda:1998mi,An:2012eh,Giunti:2007ry}. Hence the idea of extending the SMPP so as to explain the origin of neutrino mass. Different extensions of SMPP allow the neutrino to have properties such as magnetic and electric dipole moments as well as anapole moment and electric millicharge \cite{Vogel:1989iv,Giunti:2014ixa,Fukugita:2003en}. Even in the Standard Model, it is well known that the neutrinos also can have nonzero charge radius, as shown in reference \cite{Kouzakov:2017hbc}. Among these properties, the neutrino magnetic moment (NMM) has been quite studied in several research works, where different experimental constraints to this parameter were obtained, for instance, from reactor neutrino experiments \cite{Derbin:1993wy,Vidyakin:1992nf,Daraktchieva:2005kn,Deniz:2009mu,Beda:2012zz}, solar neutrinos \cite{Borexino:2017fbd,Liu:2004ny}, and astrophysical measurements \cite{Kuznetsov:2009zm,Raffelt:1999gv}. The limits achieved for the NMM are around $10^{-11}\mu_{B}$, while the prediction of the simplest extension of the Standard Model, by including right-handed neutrinos, is $3.2\times 10^{-19}\mu_{B}$ \cite{Broggini:2012df}. Furthermore, considering the representation of three active neutrinos, the magnetic moment is described by a $3\times 3$ matrix whose components are the diagonal and transition magnetic moments. A complete analysis by considering the NMM matrix and using data from solar, reactor, and accelerator experiments was presented in reference \cite{Canas:2015yoa}. In addition to NMM, the study of the remainder form factors is also important as they are a tool to probe new physics. Among them, the NEM has also been under consideration in the literature, and several constraints have been found mainly from reactor experiments and astrophysical measurements. The most restrictive bound on NEM so far, $q_{\nu}\lesssim 3.0\times 10^{-21}e$, was obtained in \cite{Raffelt:1999gv} based on the neutrality of matter. A limit of $q_{\nu}\lesssim 1.3\times 10^{-19}e$ was achieved by considering the impact of the Neutrino Star Turning mechanism on a supernova (SN) explosion~\cite{Studenikin:2012vi,Studenikin:2013yaa}. Some other astrohysical bounds come from analysis of SN 1987A neutrinos, in which effects of the galactic and extragalactic magnetic fields on the path of the neutrinos, such as $q_{\nu}\lesssim 2\times 10^{-15}e$ in \cite{Barbiellini:1987zz}. Regarding reactor neutrino experiments with ENES interactions, a bound of  $q_{\nu}\lesssim 3.7\times 10^{-12}e$  was calculated in reference \cite{Gninenko:2006fi} via data from TEXONO experiment, while one of the most strong limits was obtained by using data from GEMMA experiment and corresponds to $q_{\nu}\lesssim 1.5\times 10^{-12}e$ at $90\%$ C.L. \cite{Studenikin:2013my}. A new interaction channel was considered in the reference \cite{Chen:2014dsa}, where electron antineutrinos ($\bar{\nu}_{e}$) from reactors were studied via atomic ionizations with germanium detectors, reporting a combined limit on the neutrino charge fraction of $|\delta q|<1\times 10^{-12} e$ ($90\%$ C.L.). Most limits on the NEM from reactors experiments have been accomplished in the framework of ENES processes; however, the CENNS interactions also can be an important tool to constraint this parameter. The CENNS process had evaded the experimental detection for around forty years; nevertheless, its recent confirmation by the COHERENT collaboration renewed the interest for the study of different topics about physics beyond the standard model and CENNS \cite{Akimov:2017ade}. Some of them are NSI interactions \cite{Barranco:2005yy,Canas:2019fjw}, searches for sterile neutrinos \cite{Kosmas:2017zbh,Canas:2017umu}, measurements of precision on weak mixing angle \cite{Canas:2018rng}, probes of nuclear density distributions \cite{Patton:2012jr}, studies on neutrino magnetic moment \cite{Kosmas:2015sqa,Miranda:2019wdy}, bounds on the neutrino charge radii from analysis of COHERENT experiment data \cite{Cadeddu:2018dux} and neutrino couplings to light scalars \cite{Farzan:2018gtr}. Concerning the neutrino electric charge, recent limits of the order of $10^{-7} e$ by including neutrino flavor change, were obtained for the first time from the analysis of COHERENT experiment data~\cite{Cadeddu:2019eta}. These bounds are not competitive with those achieved for the effective charge from reactor neutrino experiments data, as mentioned in the same reference. Similar constraints on the transition charges were reported in \cite{Cadeddu:2020lky} based on data from the COHERENT experiment with argon as detection material. At the present work we will explore the sensitivity of CENNS future experiments of reactor neutrinos to neutrino electric millicharge.

\section{THE NEUTRINO ELECTRIC MILLICHARGE}

\noindent The electromagnetic interaction between a neutrino and a lepton or a quark can be expressed through a matrix element related to the electromagnetic current, which is given by the expression:
\begin{equation}
  \langle \nu(p_{f},s_{f})|J^{EM}_{\mu}|\nu(p_{i},s_{i})\rangle = i\bar{u}_{f}\Gamma_{\mu}(q)u_{i},
\label{Eq:1}  
\end{equation}
where $p_{i},p_{f},s_{i}$ and $s_{f}$ represent the initial and final neutrino momentum and spin projections, respectively \cite{Kayser:1982br,Fukugita:2003en}. The vertex function $\Gamma_{\mu}$  is described through four form factors,
\begin{eqnarray}
  \Gamma_{\mu}(q) = F_{D}(q^{2})\gamma_{\mu} + G_{D}(q^{2})(q^{2}\gamma_{\mu}-2miq_{\mu})\gamma_{5}\\
  \nonumber + M_{D}(q^{2})\sigma_{\mu\nu}q_{\nu} + E_{D}(q^{2})i\sigma_{\mu\nu}q_{\nu}\gamma_{5}.
\label{Eq:2}  
\end{eqnarray}  
Here $F_{D}(q^{2})$, $G_{D}(q^{2})$, $M_{D}(q^{2})$, and $E_{D}(q^{2})$ correspond to the charge, anapole, magnetic, and electric form factors, respectively. In the most general case, the form factors are represented by matrices including the neutrinos mass eigenstates \cite{Shrock:1982sc}. If we consider coupling with a real photon, $q^{2}=0$, then
\begin{equation}
  F_{D}(0) = q_{\nu},\;G_{D}(0) = a,\;M_{D}(0) = \mu_{\nu},\;E_{D}(0)= d,
\label{Eq:3}  
\end{equation}
being $q_{\nu},a,\mu_{\nu}$, and $d$ the neutrino millicharge, anapole moment, magnetic moment, and electric dipole moment \cite{Giunti:2008ve}. On the other hand, in the Standard Model $SU(2)_{L}\times U(1)_{Y}$ of electroweak interactions, the fact that the neutrino will be an electrically chargeless particle is a consequence of the quantization of the electric charge \cite{Giunti:2014ixa}. However, if we take into consideration the minimally extended Standard Model, which introduces a right-handed singlet neutrino $\nu_{R}$, the triangle anomaly can be removed from the $U_{B-L}$ symmetry of the SM. The introduction of an anomaly-free symmetry produces a dequantization of the electric charge; as a result, the neutrinos can be millicharged particles, whose value allowed by the measurements of non-neutrality of matter \cite{Bressi:2011yfa} and neutron charge \cite{Baumann:1988ue}  would be $Q_{\nu L}=Q_{\nu R} = \epsilon$, being $\epsilon\lesssim (-0.6\pm 3.2) \time 10^{-21}e$. Detailed information concerning the theoretical aspects of neutrino electric millicharge can be found in references \cite{Giunti:2014ixa,Fukugita:2003en} and citations therein. In the next sections we will show the phenomenological analyses employed to obtain bounds on the neutrino electric millicharge from data of ENES experiments, and from CENNS future proposals of reactor neutrinos.

\section{Bounds on NEM from ENES experiments of reactor neutrinos}

\begin{center}
\textbf{A}. $\mathbf{\chi^{2}}$ \textbf{Statistical analysis}
\end{center}
\noindent The study of the electromagnetic properties of neutrinos is usually addressed within the framework of elastic neutrino-electron scattering. In this interaction, low energy neutrinos and anti-neutrinos are scattered throughout the collision with electrons in solar, accelerator, and reactor experiments. In this analysis we take into account data from reactor experiments. The total antineutrino-electron cross section  will be composed by the Standard Model (SM) contribution, the electromagnetic (EM) term, and the interference term between SM and EM neutrino interactions, as follows, 

\begin{equation}
  \left(\frac{d\sigma}{dT_{e}}\right)^{\bar{\nu}e}_{\text{tot}} = \left(\frac{d\sigma}{dT_{e}}\right)^{\bar{\nu}e}_{\text{SM}} + \left(\frac{d\sigma}{dT_{e}}\right)^{\bar{\nu}e}_{\text{EM}} + \left(\frac{d\sigma}{dT_{e}}\right)^{\bar{\nu}e}_{\text{INT}}
\label{Eq:4}  
\end{equation}
where
\begin{eqnarray}
  \left(\frac{d\sigma}{dT_{e}}\right)^{\bar{\nu}e}_{\text{SM}}=\frac{2\text{G}^{2}_{\text{F}}m_{e}}{\pi}\left[g^{2}_{L} + g^{2}_{R}\left(1-\frac{T_{e}}{E_{\nu}}\right)^{2}\right.\\
    \nonumber - \left. g_{L}g_{R}\left(\frac{m_{e}T_{e}}{E^{2}_{\nu}}\right)\right],
  \label{Eq:5} 
\end{eqnarray}
being $\text{G}_{\text{F}}$ the Fermi coupling constant, $T_{e}$ the electron recoil energy, and $E_{\nu}$ the incident neutrino energy; while the standard coupling constants are $g_{L}=\sin^{2}\theta_{W}$, and $g_{R}=\sin^{2}\theta_{W}+1/2$ \cite{Vogel:1989iv}. Likewise, the term related to the electromagnetic component is given by
\begin{equation}
  \left(\frac{d\sigma}{dT_{e}}\right)^{\bar{\nu}e}_{\text{EM}} \simeq \frac{2\pi\alpha}{m_{e}T^{2}_{e}}q^{2}_{\nu},
\label{Eq:6}  
\end{equation}
where $\alpha$ is the fine structure constant and $q_{\nu}$ stands for the neutrino electric millicharge measured in units of electron charge \cite{Giunti:2014ixa}. Finally, the interference term was calculated based on reference \cite{Kouzakov:2017hbc}, and corresponds to:
\begin{equation}
\left(\frac{d\sigma}{dT_{e}}\right)^{\bar{\nu}e}_{\text{INT}} = \frac{2\sqrt{2\alpha}G_{F}}{T_{e}}q_{\nu}
\label{Eq:7}  
\end{equation}  
In order to calculate the total number of events (in the i-th bin), we will use the following expression:
\begin{eqnarray}
  N^{\text{th}} = \kappa\int^{E_{\nu_{\text{max}}}}_{E_{\nu_{\text{min}}}}\int^{T_{i+1}}_{T_{i}}\int^{T_{\text{max}}}_{T_{\text{min}}}\lambda(E_{\nu})\left(\frac{d\sigma}{dT_{e}}\right)^{\bar{\nu}e}_{\text{tot}}
  \label{Eq:8}\\
  \nonumber\times R(T_{e},T'_{e})dT_{e}dT'_{e}dE_{\nu}.
\end{eqnarray}
\noindent The above integral is evaluated in the real recoil energy $T_{e}$, the detected recoil energy of the electron $T'_{e}$, and the neutrino energy $E_{\nu}$. $\lambda(E_{\nu})$ represents the antineutrino spectrum from the nuclear reactor. For energies above 2 MeV, we used the $\lambda(E_{\nu})$ parametrization given by reference \cite{Mueller:2011nm}, while for $E_{\nu}<2$~MeV we employ the spectrum of electronic reactor antineutrinos showed in \cite{Kopeikin:1997ve}. Additionally, $\kappa=N_{\text{targ}}\Phi_{\bar{\nu}_{e}}t_{\text{tot}}$ corresponds to the product of the total number of targets times the total antineutrino flux times the total exposure time of the different experimental runs. Furthermore, we have used the resolution function given by the experiments,
\begin{equation}
  R(T_{e},T'_{e})=\frac{1}{\sqrt{2\pi}\sigma}\exp\left({\frac{-(T_{e}-T'_{e})^{2}}{2\sigma^{2}}}\right),
\label{Eq:9}  
\end{equation}
where $\sigma$ accounts for the error in the measurement of the kinetic energy of the electrons. For some experiments studied in this work, there is not information available related to the resolution function; accordingly, we have considered it as a delta function, namely, $R(T_{e},T'_{e})=\delta(T_{e}-T'_{e})$. Finally, we will use the $\chi^{2}$ function to carry out the statistical analysis,
\begin{equation}
  \chi^{2}=\sum^{N_{\text{bin}}}_{i=1}\frac{(N^{\text{SM}}-N^{\text{th}}(q_{\nu}))^{2}}{\Delta^{2}_{i}}.
\label{Eq:10}  
\end{equation}  
Here $N^{\text{SM}}$ represents the number of events measured by each experiment according to the Standard Model, $N^{\text{th}}$ stands for the theoretical prediction as a function of the neutrino millicharge, and $\Delta_{i}$ is the statistical error in each energy bin. ENES interactions have not been measured by the GEMMA experiment, the constraint reported in \cite{Studenikin:2013my} was obtained from estimates of the backgrounds, therefore this bound could not be derived using equation (\ref{Eq:10}). The method used in our analysis of the GEMMA experiment to obtain the respective $\chi^{2}$ profile will be shown in the part C of this section. Moreover, if a experiment reports a single energy bin, we will use the following expression,
\begin{equation}
  \chi^{2}=\frac{(N^{\text{SM}}-N^{\text{th}}(q_{\nu}))^{2}}{\sigma^{2}_{\text{stat}}},
\label{Eq:11}  
\end{equation}
where $\sigma_{\text{stat}}$ represents the statistical error.\\
\begin{center}
\textbf{B. Reactor neutrino experiments}
\end{center}

\begin{figure}[H]
 \centering
  \subfloat{
   \includegraphics[width=0.35\textwidth]{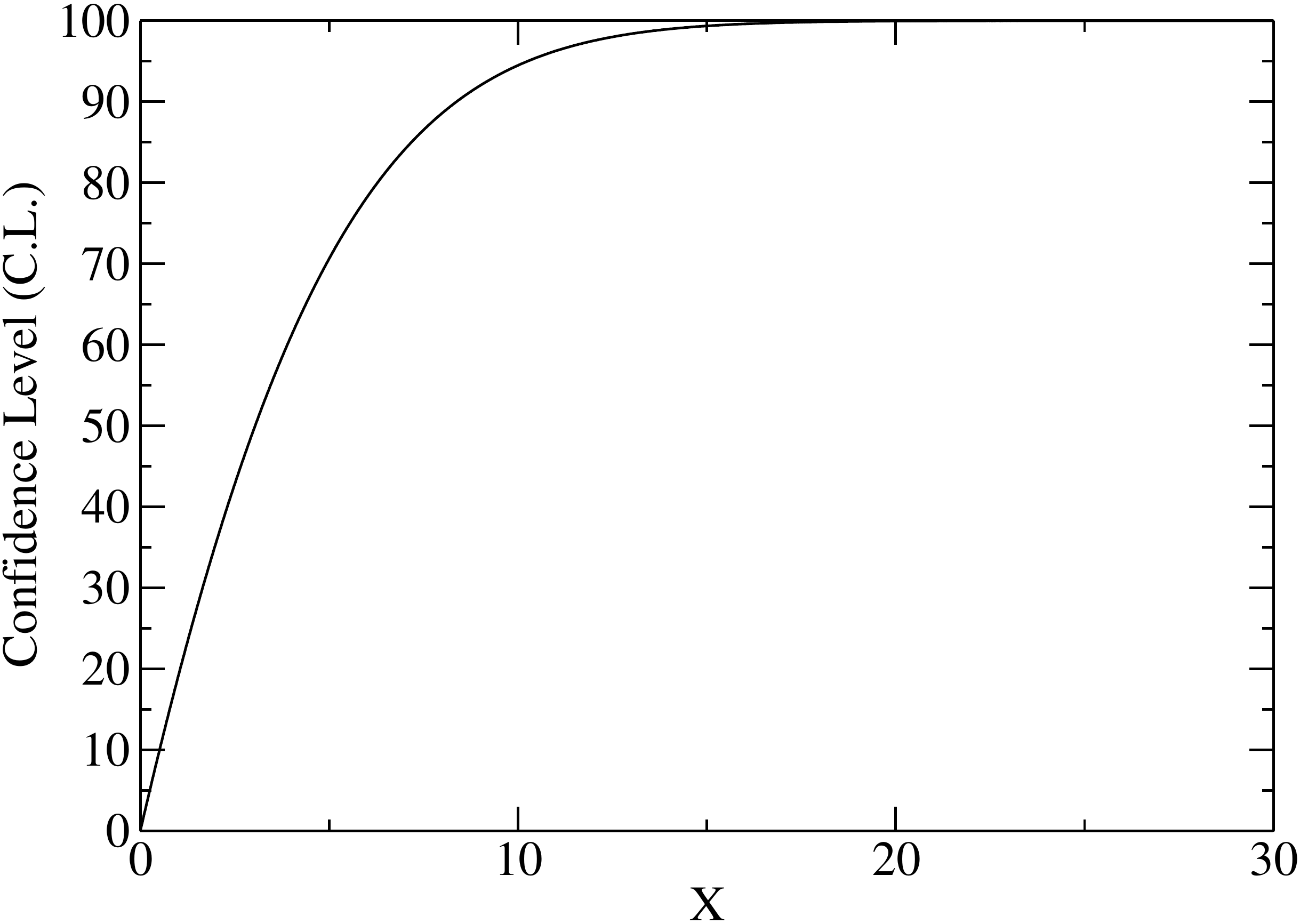}}
  \hspace{6mm}
  \subfloat{
   \includegraphics[width=0.35\textwidth]{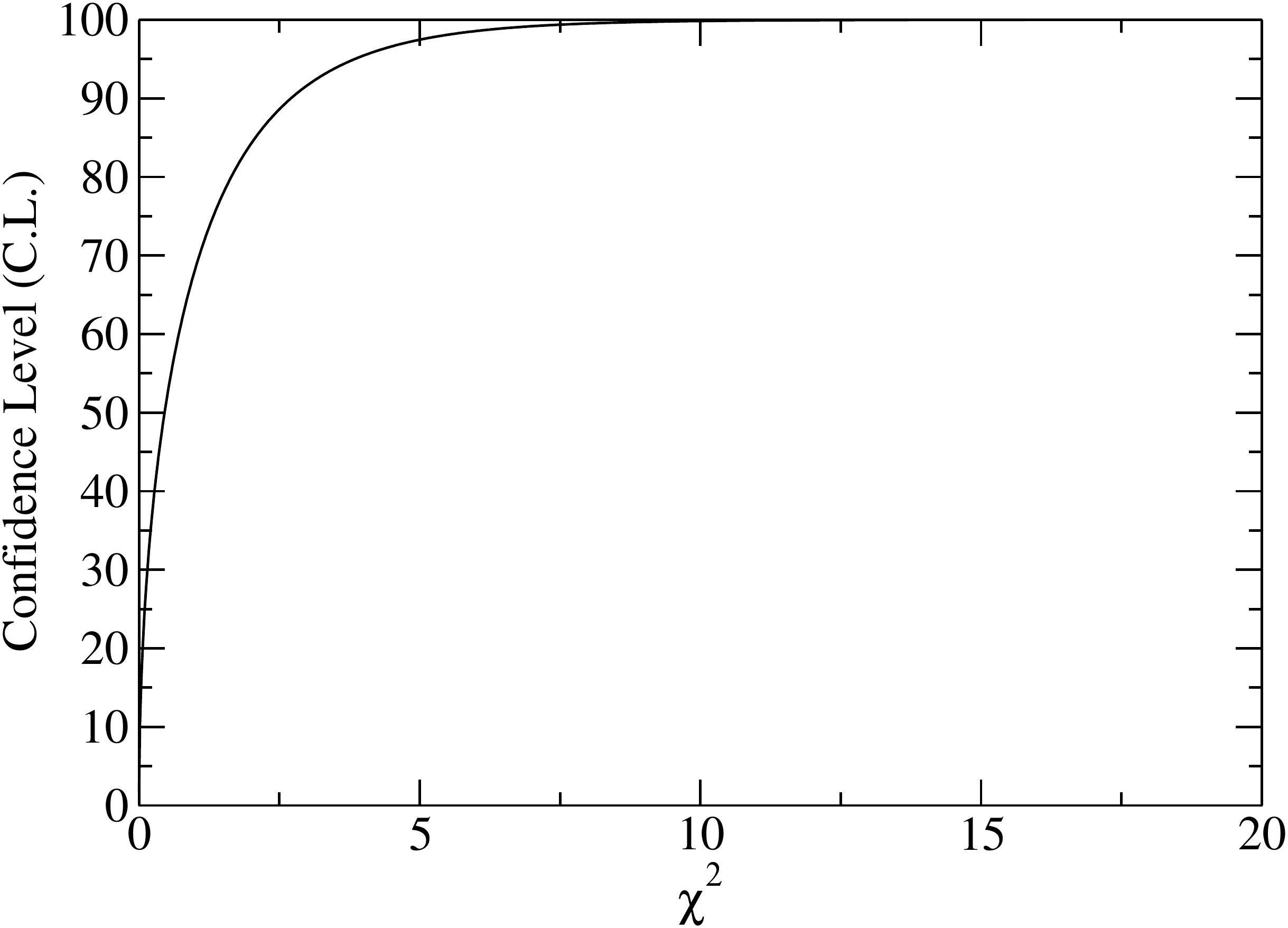}}
  \caption{\small{$\chi^{2}$ analysis for the GEMMA experiment. {\textbf{Left:} Confidence level as a function of $X$ parameter; the result is in agreement with the right panel of Figure 8 from the reference \cite{Beda:2012zz}. \textbf{Right:} Relation between the confidence level and the $\chi^{2}$ function, this result was obtained by following the Particle Data Group \cite{Tanabashi:2018oca}.}}}     
  \label{Fig:1}
\end{figure}

\noindent Regarding the NEM limits from neutrino-electron elastic processes, we used data reported by the following experiments: Rovno, Krasnosyarsk, MUNU, TEXONO, and GEMMA. And we will discribe them briefly below.
\begin{itemize}\itemsep=6pt
\item Rovno was an experiment located at the Rovno Nuclear Power Plant in Ukraine, which contained $75$~Kg of a silicon multidetector placed at $15$ m from the center of the core of the reactor. The cross section ($\bar{\nu}_{e}e$) measured in the interval $0.6-2.0$~MeV is $\sigma=(12.6\pm 0.62)\times 10^{-45}\text{cm}^{2}/\text{fission}$ \cite{Derbin:1993wy}.
\item The Krasnoyarsk experiment was developed at the nuclear power plant of Krasnoyarsk in Russia. A $103$ Kg detector of a fluoroorganic scintillator was used. The cross section for the $\bar{\nu}_{e}e$ scattering measured in the interval $3.15-5.175$~MeV corresponds to $(4.5\pm 2.4)\times 10^{-46}\text{cm}^{2}/\text{fission}$ \cite{Vidyakin:1992nf}.
\item The MUNU Collaboration has studied the existence of a magnetic moment of the electron antineutrino through ENES at low energy. The detector was located at $18$~m from the core of a $2.75$~GWth commercial nuclear reactor in the Bugey region in eastern France, being the main element of this detector a chamber filled with $\text{CF}_{4}$. The measurements were taken considering $66.6$ days live time of reactor-on data for electron recoil energies between $0.7$ and $2.0$~MeV, where $1.07\pm 0.34$ events per day was observed~\cite{Daraktchieva:2005kn}.
\item The TEXONO Collaboration employed a CsI(Tl) scintillating crystal array as detector with a total mass of $187$~Kg, placed at $28$~m from Core 1 of the Kuo-Sheng Nuclear Power Station in Taiwan. In the interval $3.0-8.0$~MeV, a total of events number of $414\pm 80\pm 61$ were registered in $624.9$ days of reactor ON and $156.1$ days of rector OFF~\cite{Deniz:2009mu}.
\item The GEMMA experiment is operating at the Kalinin Nuclear Power Plant (KNPP) in Russia. The distance between the detector of high-purity germanium (with a mass of $1.5$~Kg) and the $3\text{GW}_{\text{th}}$ reactor core is $13.9$~m \cite{Beda:2012zz}.\\\\
Moreover, currently the development of low-energy detectors is one of the main purposes of the reactor neutrino experiments. However, for increasingly lower detection threshold energies ($T<10$ KeV), the atomic effects become relevant. According to the results showed in \cite{Chen:2014dsa}, for ENES interactions, the free electron aproximation (FEA) (depicted in the equation~\ref{Eq:6}) would not be enough and effects of atomic ionization should be included. In the reference \cite{Chen:2014dsa}, the atomic effects were taken into account through a new interaction channel, $\bar{\nu}_{e}+A\to \bar{\nu}_{e}+A^{+}+e$, where the cross section was calculated by using the Multi-Configuration Relativistic Random-Phase Approximation (MCRRPA) \cite{Chen:2013lba}. The results display an enhancement of an order of magnitude in the differential cross section ($\bar{\nu}_{e}-A(\delta_{q})$) over FEA, when atomic ionization effects are considered. The analysis allowed to obtain a neutrino charge fraction of $|\delta q|<1\times 10^{-12}$ ($90\%$ C.L.) by including combined data from GEMMA and TEXONO experiments \cite{Chen:2014dsa}. As a first approximation, in our calculations, we will not include atomic effects, taking into account that our priority is to initially show the advantages of CENNS interactions over ENES processes in constraining of the neutrino electric millicharge.

\end{itemize}
\vspace{0.05cm}
\begin{center}
\textbf{C. Limits on neutrino electric millicharge}
\end{center}
In the different calculations corresponding to the bounds on NEM, we will use the $\chi^{2}$ analysis given by equation (\ref{Eq:10}) or (\ref{Eq:11}). However, as we mentioned above, in the case of GEMMA experiment we will use a different method to obtain the $\chi^{2}$ profile.
As shown in \cite{Studenikin:2013my}, the ratio between the electromagnetic cross sections for the NEM and NMM contributions is given by,
\begin{equation}  
  R=\frac{\left(\frac{d\sigma}{dT}\right)_{q_{\nu}}}{\left(\frac{d\sigma}{dT}\right)_{\mu_{\nu}}}=\frac{2m_{e}}{T}\left(\frac{q_{\nu}}{\mu_{\nu}}\right)^{2}\left(\frac{\mu_{B}}{e}\right)^{2}\lesssim 1, 
\label{Eq:12}
\end{equation}
where it has been considered that NEM ($q_{\nu}$) is smaller than NMM ($\mu_{\nu}$), $\mu_{B}$ and $e$ represent the Bohr magnet and the fundamental charge, respectively. From equation \ref{Eq:12} it follows that,
\begin{equation}
  q^{2}_{\nu}\lesssim \frac{T_{e}}{2m_{e}}\left(\frac{\mu_{\nu}}{\mu_{B}}\right)^{2}e^{2}.
\label{Eq:13}  
\end{equation}
Finally, the limit on NEM was computed by replacing the values: $\mu_{\nu}<2.9\times 10^{-11}\mu_{B}$ and $T_{e}=2.8$ KeV in the equation \ref{Eq:13}, as presented in \cite{Studenikin:2013my}. Nevertheless, we need the $\chi^{2}$ profile as a function of $q_{\nu}$ from the GEMMA experiment for combining it with the results from the rest of the reactor neutrino experiments under review. Therefore, and considering that the calculation to get the bound on NEM ($q_{\nu}\lesssim 1.5\times 10^{-12}e$) was based on \cite{Beda:2012zz}, we will employ the parametrization presented there in order to express the electromagnetic cross section for NMM (equation (3.3) of \cite{Beda:2012zz}),
\begin{equation}
  \left(\frac{d\sigma}{dT_{e}}\right)_{\text{EM}} = C\left(\frac{1}{T_{e}}-\frac{1}{E_{\nu}}\right)*X,
\label{Eq:14}  
\end{equation}
being $C$ a constant and $X=\left(\frac{\mu_{\nu}}{10^{-11}\mu_{B}}\right)^{2}$. Furthermore the probability density function is given by,
\begin{equation}
  f(X) = \frac{1}{\sqrt{2\pi}\sigma}\exp\left[\left(-\frac{(X-X_{0})^{2}}{2\sigma^{2}}\right)\right],
\label{Eq:15}  
\end{equation}
where $\sigma=6.9$, and $X_{0}=-5.79$. Making use of the last expression, we found the values for the confidence level as function of the $X$ parameter. The result can be seen in the left panel of Figure \ref{Fig:1}, which corresponds to the right panel of Figure 8 from the reference \cite{Beda:2012zz}.

\noindent As we saw above, the NMM is a function of $X$ ($\mu_{\nu}=\sqrt{X}~10^{-11}\mu_{B}$), then the limit on the neutrino millicharge is given by,
\begin{equation}
  q_{\nu}\lesssim \sqrt{X}\left(\frac{T}{2m_{e}}\right)^{1/2}10^{-11}e.
\label{Eq:16}  
\end{equation}
\noindent On the other hand, the values of $\chi^{2}$ and the $X$ parameter are related through the composite function: $\chi^{2}(\text{CL}(X))$, and the relation between the function $\chi^{2}$ and the CL can be found following a recomendation of the Particle Data Group \cite{Tanabashi:2018oca}. Considering the goodness-of-fit statistics is described by a $\chi^{2}$ probability distribution function (p.d.f) \cite{Tanabashi:2018oca}, the $p$-value corresponds to
\begin{equation}
  p=\int^{30}_{\chi^{2}}f(z,n_{\text{d}})dz,
\label{Eq:17}  
\end{equation}  
where the number of degrees of freedom is $n_{d}=1$, and $f(z,n_{\text{d}})$ is the $\chi^{2}$ p.d.f, 
\begin{equation}
  f(z,n_{\text{d}})=\frac{1}{\sqrt{2\pi z}}e^{-z/2}.
\label{Eq:18}  
\end{equation}
The right panel  of the Figure~\ref{Fig:1} shows the result for the confidence level as a function of $\chi^{2}$, obtained by using the above method from PDG. Finally, from $\chi^{2}(\text{CL}(X))$ and the equation \ref{Eq:16}, we reproduced the limit $q_{\nu}\lesssim 1.5\times 10^{-12}e$ ($90\%$ C.L.) obtained in \cite{Studenikin:2013my} based on the use of GEMMA experiment data. The corresponding $\chi^{2}$ profile will be included in our combined analysis.
\onecolumngrid

\twocolumngrid

\begin{figure}[h]
  \includegraphics[width=0.45\textwidth]{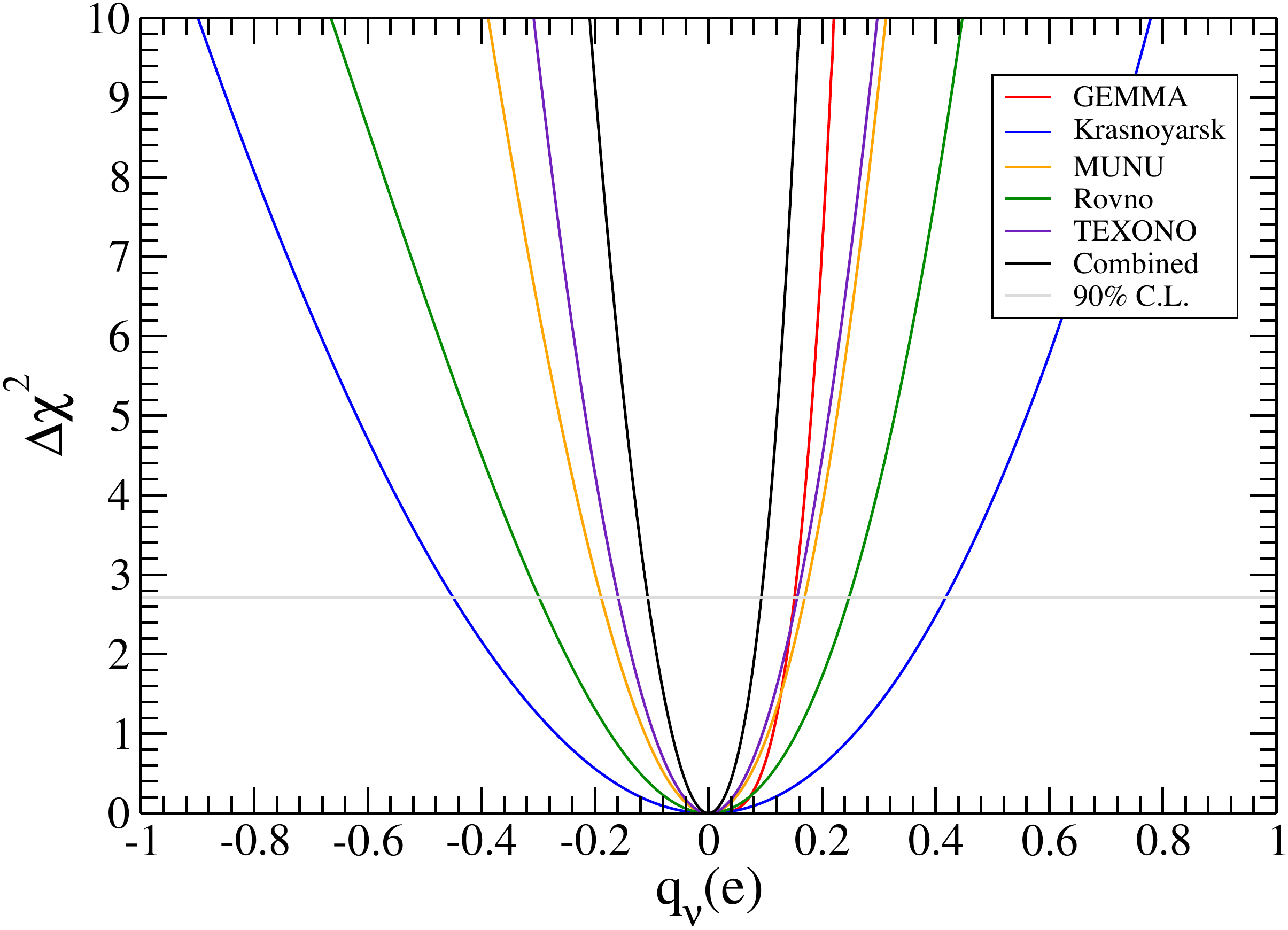}
\caption{\small{$\Delta\chi^{2}$ sensitivity profile for the neutrino charge, $q_{\nu}$ in units of $10^{-11} e$, obtained from data of reactor neutrino experiments. The black line corresponds to the combined profile.}}
\label{Fig:2}    
\end{figure}

\noindent The results for the $\Delta\chi^{2}$ sensitive profile (in terms of NEM) for each experiment, together with the combined analysis, are displayed in Figure \ref{Fig:2}. The Table \ref{Tab:1} shows the values for the bounds on the NEM, the limits from the combination of all experiments are $-1.1\times 10^{-12}e < q_{\nu} < 9.3\times 10^{-13}e$ ($90\%$ C.L.). In this result the upper constraint is lower than each of the individual bounds obtained from the other experiments studied here, including the limit achieved in \cite{Studenikin:2013my} based on the analysis of GEMMA experiment data ($q_{\nu} < 1.5\times 10^{-12}e$). According to the results in Table \ref{Tab:1}, the most important contributions to the combined limit are given by GEMMA, TEXONO and MUNU experiments, respectively.

\begin{table}[H]
\begin{center}
\begin{tabular}{|c|c|}
\hline
 \textbf{EXPERIMENT} & \textbf{LIMIT}\\
\hline
Rovno & $-3.0 < q_{\nu} < 2.5$\\
\hline
Krasnoyarsk & $-4.5 < q_{\nu} < 4.2 $\\
\hline
MUNU & $-1.9  < q_{\nu} < 1.7 $\\
\hline
TEXONO & $-1.6 < q_{\nu} < 1.6 $\\
\hline
GEMMA & $q_{\nu} < 1.5 $\\
\hline
\textbf{Combined} & $\mathbf{-1.1 < q_{\nu} < 0.93}$\\
\hline
\end{tabular}
\caption{\small{90\% C.L. Limits on the neutrino millicharge, $q_{\nu}$ in units of $10^{-12}e$, obtained from data of reactor neutrino experiments.}}
\label{Tab:1}
\end{center}
\end{table}

\section{Bounds on NEM from CENNS future experiments of reactor neutrinos}
\begin{center}
\textbf{A}. \textbf{CENNS Interactions}
\end{center}

\noindent In 1974, Daniel Freedman suggested the possibility of neutrinos interact coherently with nucleus, based on the discovery of the weak neutral current in CERN and NAL \cite{Freedman:1973yd}.
This coherence condition takes place as long as the momentum transfer $Q\lesssim1/R$ ($R$ is the nuclear radius), which is satisfied by neutrinos with energies up to $\sim 50$ MeV. The discovery of CENNS was reported by the COHERENT collaboration in 2017, and the observation corresponds to a signal at a $6.7\sigma$ confidence level by using a $14.6$ Kg CsI[Na] target as a detector of neutrinos from the Spallation Neutron Source (SNS) at Oak Ridge National Laboratory \cite{Akimov:2017ade}. In this case, the total cross section of the coherent elastic neutrino-nucleus scattering is given by,

\begin{equation}
 \left(\frac{d\sigma}{dT}\right)^{\text{coh}}_{\text{tot}} = \left(\frac{d\sigma}{dT}\right)^{\text{coh}}_{\text{SM}} + \left(\frac{d\sigma}{dT}\right)^{\text{coh}}_{\text{EM}} + \left(\frac{d\sigma}{dT}\right)^{\text{coh}}_{\text{INT}}
\label{Eq:19}   
\end{equation}  
where the contribution of the Standard Model is as follows,
\begin{equation}
  \left(\frac{d\sigma}{dT}\right)^{\text{coh}}_{\text{SM}} = \frac{G^{2}_{\text{F}}M}{\pi}\left[1-\frac{MT}{2 E^{2}_{\nu}}\right]\left(g^{p}_{V}Z + g^{n}_{V}N\right)^{2}F(Q^{2}).
\label{Eq:20}    
\end{equation}
\\
\noindent Here $G_{\text{F}}$ is the Fermi coupling constant, $M$ and $T$ represent the mass and the recoil energy of the nucleus, respectively; $Z$ and $N$ correspond to the number of protons and neutrons contained in the nucleus \cite{Barranco:2005yy}. Likewise, $F(Q^{2})=F_{Z}(Q^{2})=F_{N}(Q^{2})$ stands for the nuclear form factor, and the vector couplings of the neutral current between neutrinos and protons (neutrons) are described by
\begin{eqnarray}
\nonumber  
g^{p}_{V} &=& \rho^{NC}_{\nu N}\left(\frac{1}{2}-2\hat{\kappa}_{\nu N}\hat{S}^{2}_{Z}\right) + 2\lambda^{uL} + 2\lambda^{uR} + \lambda^{dL} + \lambda^{dR},\\
g^{n}_{V} &=& -\frac{1}{2}\rho^{NC}_{\nu N} + \lambda^{uL} + \lambda^{uR} + 2\lambda^{dL} + 2\lambda^{dR}.
\label{Eq:21}    
\end{eqnarray}  
The radiative corrections in the last expressions are taken from the PDG \cite{Tanabashi:2018oca}. As mentioned above, the transfer momentum $Q^{2}$ must be sufficiently small in order to satisfy the coherence condition. Thus, in the limit $Q^{2}\to 0$, we have the approximation $F(Q^{2})\simeq 1$, which means a full coherence \cite{Kerman:2016jqp}. Based on \cite{Giunti:2015gga}, the electromagnetic contribution and the interference term are given by,
\begin{equation}
  \left(\frac{d\sigma}{dT}\right)^{\text{coh}}_{\text{EM}} = \frac{2\pi Z^{2}}{MT^{2}}\left(1 - \frac{MT}{2E^{2}_{\nu}}\right)q^{2}_{\nu},
\label{Eq:22}    
\end{equation}
and
\begin{equation}
  \left(\frac{d\sigma}{dT}\right)^{\text{coh}}_{\text{INT}} = \frac{\sqrt{8}G_{F}C_{V}Z}{T}\left(1 - \frac{MT}{2E^{2}_{\nu}}\right)q_{\nu},
  \label{Eq:23}    
\end{equation}  
respectively; where $C_{V} = g^{p}_{V}Z + g^{n}_{V}N$ and $q_{\nu}$ represents the neutrino charge in units of the electron charge, $q_{\nu}=e_{\nu}e$. For the purpose of calculating the events number in a reactor neutrino detector, we used an expression similar to Equation (\ref{Eq:8}), taking into consideration the corresponding total CENNS cross section. Likewise, to obtain the constraints to the neutrino millicharge, we carried out an $\chi^{2}$ analysis like the one described in Equation (\ref{Eq:10}), by including the respective expression for the number of events in CENNS interactions.\\
\begin{center}
\textbf{B. Future neutrino reactor experiments with CENNS}
\end{center}
\noindent Several experimental proposals of reactor neutrinos that include CENNS  interactions are being developed around the world. Each detector involves different features in terms of technology, configuration, and system detection. In our analysis, we have considered some experiments whose expectation is to achieve results in the near future, such as TEXONO, CONNIE, CONUS, RED100, and MINER. A short description of each of these experiments is shown below.
\begin{itemize}\itemsep=6pt
\item The TEXONO Collaboration. The experiment is located at the Kuo-Sheng Neutrino Laboratory (Taiwan), where a distance of $28$ m separates the detector from the $2.9$ GW reactor core. Among the current goals of this collaboration is the development of germanium detectors with target mass of $1$ Kg, with sensitivities of up $100$ eV and low-background specifications \cite{Wong:2010zzc}. For the purposes of our study we will consider a quenching factor $Q_{f}=1$, and an antineutrino flux of $\Phi_{\bar{\nu}_{e}}=1\times 10^{13} \text{cm}^{-2}\,\text{s}^{-1}$; the estimation for the expected number of events is $27962$. \cite{Kosmas:2015vsa}.
\item The CONNIE Experiment. This detector is located at a distance of $30$ m from the core of the $3.8 \text{GW}_{th}$ nuclear reactor in the Almirante Alvaro Alberto Nuclear Power Plant, in Rio de Janeiro, Brazil. The core of the detector contains an array of Charged-Coupled Devices (CCDs), with silicon as detection material. The neutrino flux in the detector is $\Phi_{\bar{\nu}_{e}}=7.8\times 10^{12}\text{cm}^{-1}\text{s}^{-1}$, and the expected rate of events would be $16.1\;\text{evt}\,\text{kg}^{-1}\,\text{day}^{-1}$ \cite{Aguilar-Arevalo:2016qen}. In an exposition of $1$ kg-year, the total number of expected events corresponds to $5877$, considering threshold energies up to $28$ eV \cite{Aguilar-Arevalo:2016khx}.
\item The CONUS Experiment is being developed in a nuclear power plant in Brokdorf, Germany, where the detector is placed $17$ m from the core of the nuclear reactor. In the first measurements a mass of $4$ Kg Ge was condidered, and data collection is estimated to last five years. The total antineutrino flux at the detector is $\Phi_{\bar{\nu}_{e}}=2.5\times 10^{13} \text{cm}^{-2}\,\text{s}^{-1}$, and the threshold energy is $0.1$ KeV. The expected number of the events under these conditions is $31200\,\text{evt}\,\text{year}^{-1}$\cite{Farzan:2018gtr,Lindner:2016wff}.
\item The RED100 Experiment. This detector is located at the Kalinin Nuclear Power Plant at $19$m from the reactor core. The expectation is taking data by using a fiducial mass of $100$ Kg of $136$-Xe. The antineutrino flux estimated at the detector location is $\Phi_{\bar{\nu}_{e}}=1.35\times 10^{13}\text{cm}^{-2}\text{s}^{-1}$. In our analysis we included the most conservative estimation for the number of the expected events ($1020\,\text{events}\,\text{day}^{-1}$), and the threshold energy corresponds to $500$ eV \cite{Akimov:2012aya,Akimov:2017hee}.
\item The MINER Experiment. The proposal of the Mitchell Institute Neutrino Experiments at Reactor experiment is to use a combination of $^{72}\text{Ge}$ and $^{28}\text{Si}$ detectors at $2:1$ mass ratio, with a total mass of $30$ Kg. The mean distance between the detector and the reactor core is $1$ m. The anti-neutrino flux estimated at this location is $\Phi_{\bar{\nu}_{e}}=2.5\times 10^{13} \text{cm}^{-2}\text{s}^{-1}$. The configuration allows for reaching a threshold energy of $10$ eV, and $5-20\,\text{events}\,\text{Kg}^{-1}\,\text{day}^{-1}$ are expected at the detector. In the calculation we have taken an events rate of $5~\text{events}\,\text{Kg}^{-1}\,\text{day}^{-1}$ \cite{Dutta:2015vwa,Agnolet:2016zir}.
\end{itemize}  
\vspace{0.05cm}
\begin{center}
\textbf{C. $\chi^{2}$ statistical analysis and limits on NEM}
\end{center}

\begin{figure}[H]    
  \hspace{0.2cm}
   \includegraphics[width=0.45\textwidth]{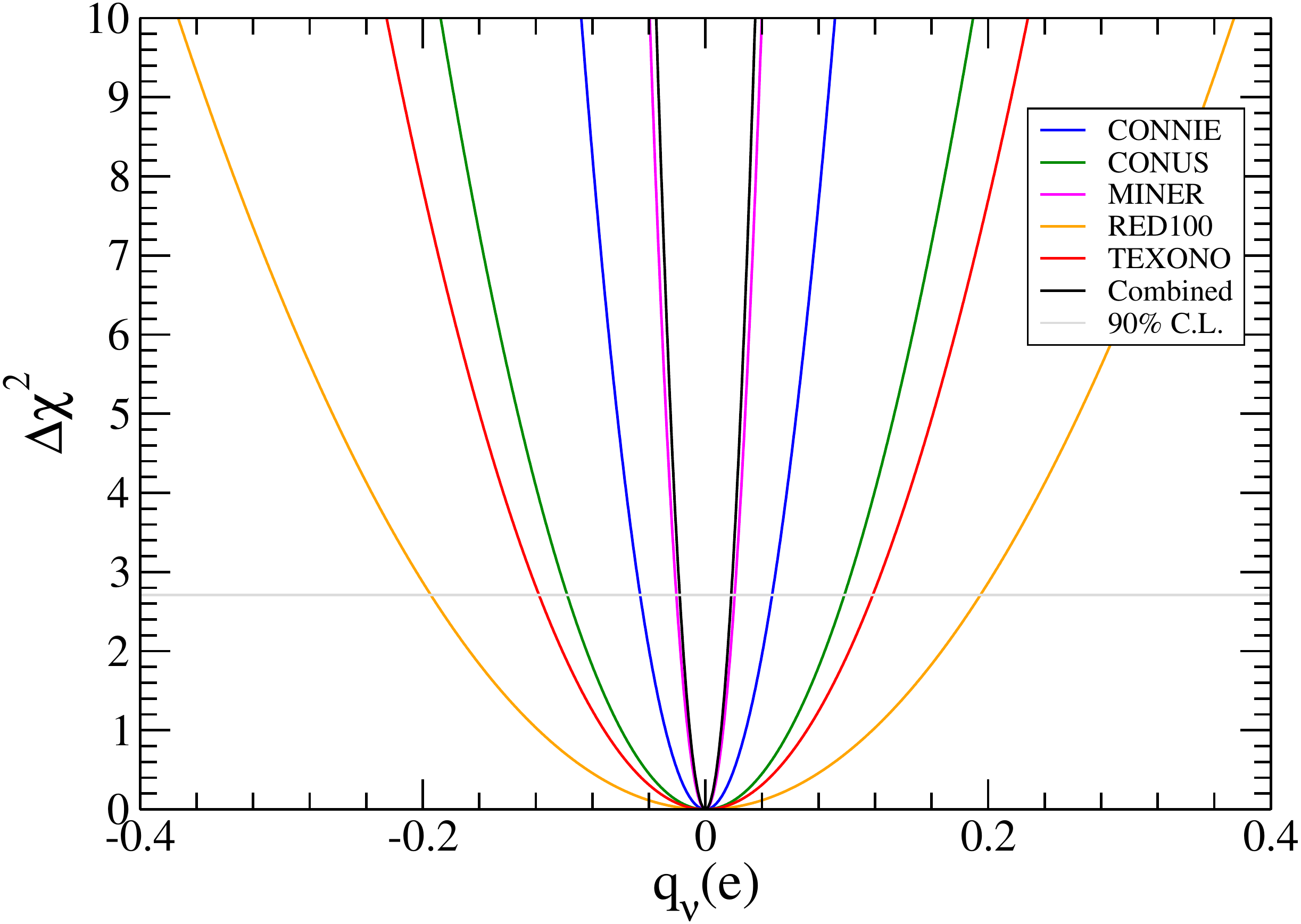}
   \caption{\small{$\Delta\chi^{2}$ sensitivity profile as a function of the neutrino charge, $q_{\nu}$ in units of $10^{-12}e$, achieved from data of CENNS future proposals of reactor neutrinos. The results were obtained by considering only statistical uncertainties in the event spectrum.}}
\label{Fig:3}    
 \end{figure}

\noindent As we mentioned above, in the framework of CENNS interactions, our main purpose is to determine the responsiveness of future experiments of reactor neutrinos to the NEM. In a first approximation we assume only future statistical errors in the event spectrum at the detector; nevertheless, we will also include possible future systematic uncertainties in a second assessment. In this last case, we will choose the parametrization given by $\sigma_{\text{syst}} = pN^{\text{th}}$, where $p$ represents the percentage of the systematic uncertainty. Then, the expression for the $\chi^{2}$ calculation is as follows, 
\begin{equation}
  \chi^{2}=\frac{(N^{\text{SM}}-N^{\text{th}}(q_{\nu}))^{2}}{\sigma^{2}_{\text{stat}} + \sigma^{2}_{\text{syst}}},
\label{Eq:24}  
\end{equation}
where $N^{\text{SM}}$ corresponds to the events number that each experiment expects to measure according to the SM, and $N^{\text{th}}$ represents the theoretical number of events as a function of NEM, while $\sigma_{\text{stat}}$ denotes the statistical error. The results from the first scenario (only statistical uncertainties) are shown in Table \ref{Tab:2} and Figure \ref{Fig:3}. In this case, the limits $-1.8\times 10^{-14}e < q_{\nu} < 1.8\times 10^{-14}e$ ($90\%$ C.L.) were obtained for the combination of the experiments under study. As we can see, these last constraints are roughly two orders of magnitud lower than the combined bounds derived from ENES experiments, which would represent an outstanding result, taking into account that it will come from terrestrial experiments.

\begin{table}[H]
  \begin{center}
\resizebox{6cm}{!}{    
\begin{tabular}{|c|c|}
\hline
\; \textbf{EXPERIMENT}\; & \textbf{LIMIT}\\
\hline
CONNIE & $-4.6  < q_{\nu} <  4.7$\\
\hline
CONUS & $-9.8  < q_{\nu} <  9.8 $\\
\hline
MINER & $-2.0 < q_{\nu} <  2.1 $\\
\hline
RED100 & $-19 < q_{\nu} <  19$\\
\hline
TEXONO & $-12 < q_{\nu} <  12$\\
\hline
\textbf{Combined} & $\mathbf{-1.8 < q_{\nu} <  1.8}$\\
\hline
\end{tabular}}
\caption{\small{90\% C.L. bounds on the neutrino millicharge, $q_{\nu}$ in units of $10^{-14} e$, from data of different CENNS future experiments of reactor neutrinos. Here only statistical errors in the event spectrum were included.}}
\label{Tab:2}
 \end{center}
  \end{table}  
\noindent 
 
\begin{figure}[h]
 \centering
 \subfloat{
    {\small (a)} 
    \includegraphics[width=0.43\textwidth]{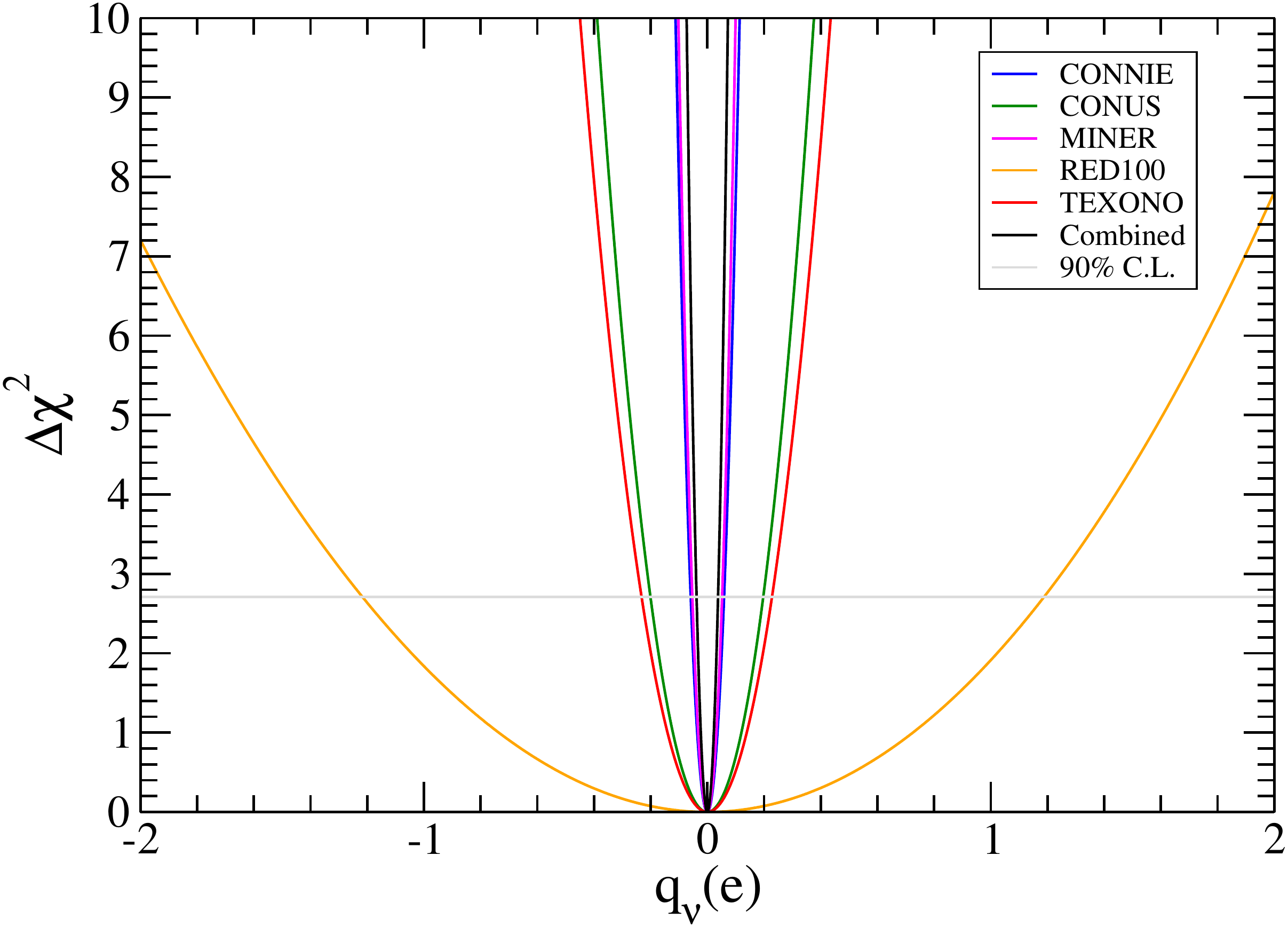}}\\ 
 \subfloat{
    {\small (b)} 
    \includegraphics[width=0.43\textwidth]{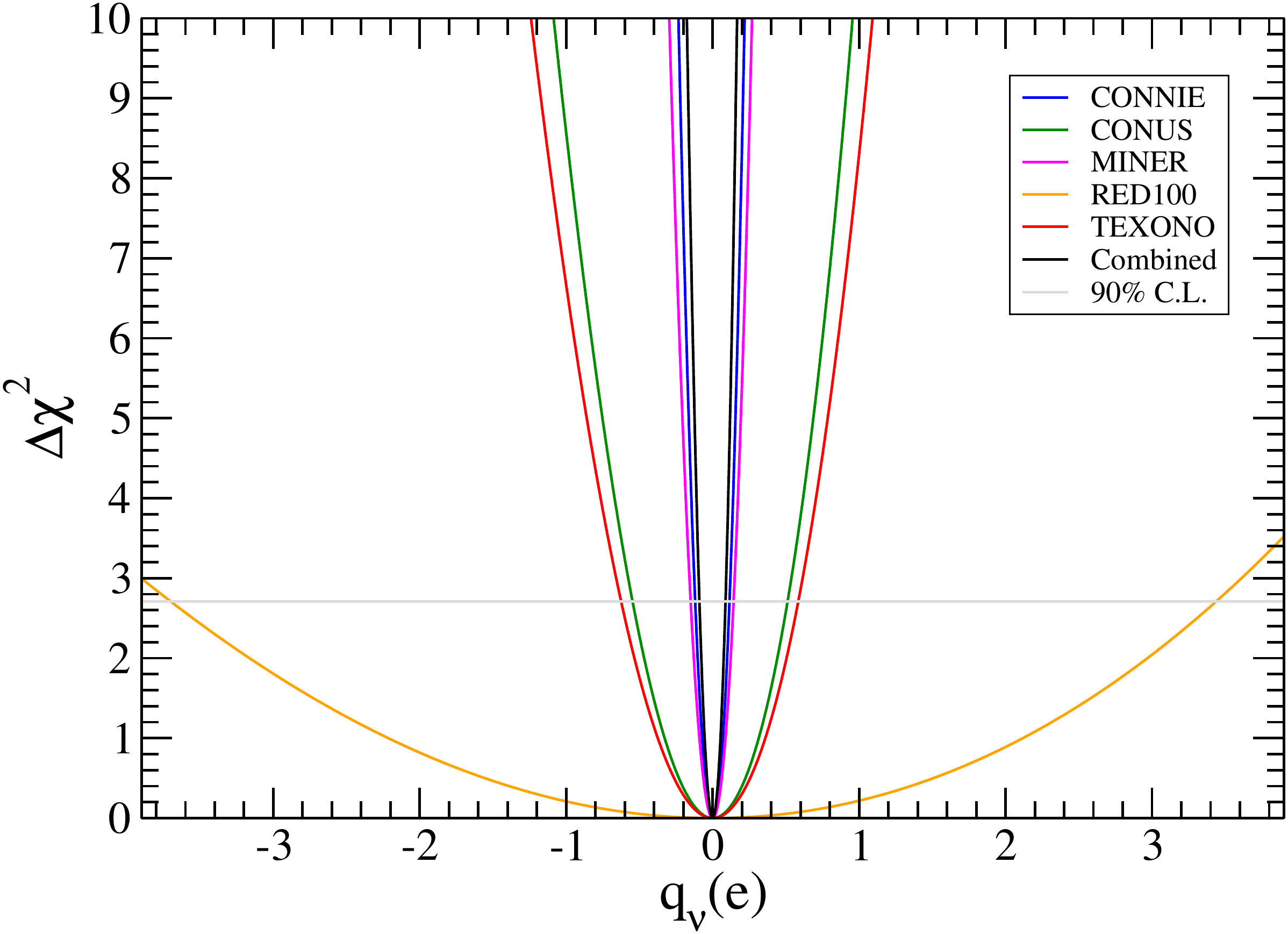}}\\
    \caption{\small{$\Delta\chi^{2}$ profile for the neutrino charge, $q_{\nu}$ in units of $10^{-12} e$, from data of CENNS future experiments of reactor neutrinos, including systematic uncertainties. The top panel (a) shows the results with the $1\% N^{\text{th}}$ of systematic error and the bottom panel (b) corresponds to the $3\% N^{\text{th}}$ of systematic uncertainty.}}
    \label{Fig:4}
\end{figure}

 \noindent Among the lower constraints, the bounds obtained from the MINER, CONNIE, and CONUS experimental proposals stand up from the rest of the experiments studied here. As we mentioned above, in the case of the MINER proposal that produces the strongest individual limits, the detection material will be composed by two different elements $^{72}$Ge and $^{28}$Si, which could indicate that a combination of different materials in the detector might be a favorable option to improve the sensitivity to the neutrino electric millicharge in CENNS experiments. The constraints achieved by considering systematic errors are displayed in Table \ref{Tab:3} and Figure \ref{Fig:4}. In this calculation, we included two values for the systematic uncertainty ($p=1\%$ and $p=3\%$) in the theoretical number of events. It is worth mentioning that even for a systematic error of the order of $3\%$ of the number of theoretical events, the combined limits are still of the order of $10^{-14}e$. This result could indicate that as the experiments reduce the systematic uncertainties, lower and lower constraints can be obtained on the neutrino electric millicharge.
 
 \begin{table}[H]
   \centering
\small{     
\begin{tabular}{|c|c|c|}
\hline
 \textbf{\,EXPERIMENT\,} & \; \textbf{LIMIT($p=1\%$)}\; & \;\textbf{LIMIT($p=3\%$)}\;\\
\hline
CONNIE & $-5.9 < q_{\nu} < 5.9$ & $-12 < q_{\nu} < 12$\\
\hline
CONUS & $-20 < q_{\nu} < 20$ & $-55 < q_{\nu} < 51$\\
\hline
MINER & $-5.3 < q_{\nu} < 5.2$ &  $-15 < q_{\nu} < 14$\\
\hline
RED100 &  $-120 < q_{\nu} < 120$ &  $-370 < q_{\nu} < 340$\\
\hline
TEXONO & $-23 < q_{\nu} < 23$ & $-63 < q_{\nu} < 59$ \\
\hline
\textbf{Combined} & $\mathbf{-3.8 < q_{\nu} < 3.8}$ & $\mathbf{-9.0 < q_{\nu} < 8.8}$  \\
\hline
\end{tabular}
}
\caption{\small{90\% C.L. Constraints on the neutrino charge, $q_{\nu}$ in units of $10^{-14}e$, obtained from several CENNS experimental proposals of reactor neutrinos. Both statistical and systematic errors in the event spectrum were included.}}
\label{Tab:3}
 \end{table}
 
\section{CONCLUSIONS}

\noindent There is an increasing interest in knowing more about the electromagnetic neutrino properties, as the neutrino electric millicharge. Different research works have been focused on obtaining limits for this parameter, including experimental and theoretical considerations. As regards to ENES processes, one of the most important limits, ${q_{\nu}\lesssim 1.5\times 10^{-12}}e$ ($90\%$ C.L.), was reported in \cite{Studenikin:2013my} from the analysis of GEMMA experiment data. In relation to CENNS interactions, limits of the order of $10^{-7}e$ on the neutrino transition charges were obtained in \cite{Cadeddu:2019eta,Cadeddu:2020lky} by using data for the first time from the COHERENT experiment. In this work, first we performed statistical analyses by considering data from past and current ENES neutrino experiments to restrict NEM. Then, and taking advantage of the great potential of coherent elastic neutrino-nucleus scattering, we studied the possibilities of some CENNS future experiments of reactor neutrinos for setting bounds on the neutrino electric millicharge. The combined analyses were made by including data and estimations from different experimental collaborations. In the context of ENES experiments, we obtained the combined limits $-1.1\times 10^{-12}e < q_{\nu} < 9.3\times 10^{-13}e$, being this upper limit lower than the bound reported in \cite{Studenikin:2013my} from the use of GEMMA experiment data. This indicates that the combination of data from different experiments gives rise to a more restrictive constraint than each limit achieved from the individual experiments. As regards CENNS future proposals, in a first aproximation and considering only possible statistical errors in the calculations, the limits achieved here are $-1.8\times 10^{-14}e < q_{\nu} < 1.8\times 10^{-14}e$ at $90\%$ C.L. Additionally, we computed the constraints by including probable systematic uncertainties corresponding to the $1\%$ and $3\%$ in the event spectrum at the detector, obtaining the limits: $-3.8\times 10^{-14}e < q_{\nu} < 3.8\times 10^{-14}e$ and $-9.0\times 10^{-14}e < q_{\nu} < 8.8\times 10^{-14}e$ at $90\%$ C.L., respectively. According to these results, CENNS combined constraints are approximately two orders of magnitude lower than those derived from ENES experiments. Which could indicate that in the near future, the CENNS experiments of reactor neutrinos will be one of the most important option to improve the limits on the neutrino electric millicharge, as the sensitivity increases at the detectors. In spite of that the bounds on NEM from CENNS proposals are larger than those coming from astrophysical measurements \cite{Raffelt:1999gv}, this difference could be further reduced with the technological advancements in CENNS detectors.

\begin{center}
\textbf{\small{DATA AVAILABILITY}}
\end{center}
\noindent The data used to support the findings of this study are included within the submitted article. The data corresponding to the elastic neutrino-electron scattering experiments are shown in section III. B, and the data from the coherent elastic neutrino-nucleus scattering experiments were included in section IV. B. These data are publicly available for each of the experiments considered in this work. The references and other details are given in the manuscript.\\

\begin{center}
\textbf{\small{CONFLICTS OF INTEREST}}
\end{center}
\noindent The authors declare that they have no conflicts of interest.
\acknowledgments
     {\noindent This work was supported by Universidad Santiago de Cali (USC) under grant 935-621120-G01, and by the CONACYT grant A1-S-23238 (Mexico). The author would like to thank O.G. Miranda for his important comments to improve this work.}

\end{document}